\documentstyle[sprocl,epsfig]{article}

\def\Journal#1#2#3#4{{#1} {\bf #2}, #3 (#4)}

\def\NPB{{\em Nucl. Phys.} B}
\def\PLB{{\em Phys. Lett.}  B}
\def\PRL{\em Phys. Rev. Lett.}
\def\PRD{{\em Phys. Rev.} D}

\begin{document}

\title{THE AXINO-GRAVITINO COSMOLOGY}

\author{ JIHN E. KIM }

\address{Lyman Laboratory of Physics, Harvard University,
\\ MA 02138, USA, and\\
Department of Physics, Seoul National University,\\
Seoul 151-742, Korea \footnote{Permanent address}}


\maketitle\abstracts{
A cosmological scenario of a light axino and a lighter gravitino
is presented \cite{ckk,kk}. The most important consequence is that it can
mimick the mixed dark matter (MDM) model of the large scale
structure formation. The presence of axino and gravitino is inevitable
in supergravity extension of the invisible axion solution of the
strong CP problem. The possibility of a light gravitino is popular
in the recent efforts of gauge mediated supersymmetry breaking. }
  
\section{Introduction}

The minimal standard model contains 19 free parameters. Among
these the parameters $\theta_{\rm QCD}$ and the Higgs boson mass
are of most fundamental importance which have led to

\indent ---strong CP problem and axions,

\indent ---gauge hierarchy problem, supersymmetry, etc.

\noindent The parameter problem is generally requiring an
understanding of WHY the parameter takes the value chosen
by nature.

Here I will discuss one scenario for the dark matter arising
from the attempt to solve the above parameter problems:
axion, supersymmetry, axino and gravitino.

\section{The $\theta_{\rm QCD}$ Parameter Problem}

Quantum chromodynamics before 1975 was described by
\begin{equation}
{\cal L}=-{1\over 2g^2}{\rm Tr}F_{\mu\nu}F^{\mu\nu}
+\bar q(iD^\mu\gamma_\mu-M)q.
\end{equation}
After 1975, it has been known to be inevitable to include,
due to the discovery of the instanton configuration~\cite{bpt}, 
\begin{equation}
+{\bar\theta\over 16\pi^2}{\rm Tr}F_{\mu\nu}\tilde F^{\mu\nu}.
\end{equation}
It must be included if there is no massless quark. This $\bar\theta$
term is a total divergence, but the gauge field configuration at
infinity gives a nontrivial surface term. Thus $\bar\theta$ parameter
is physical and strong interactions violate the CP symmetry. 
>From the upper bound of neutron electric dipole moment
$|d_n|_{\rm exp}<1.2\times 10^{-25} e$cm~\cite{altarev}, we obtain
a bound on $\bar\theta$
\begin{equation}
|\bar\theta|<10^{-9}.
\end{equation}

The essence of the strong CP problem is, $\lq\lq$Why is $\bar\theta$
so small?" The most attractive solution of the strong CP problem is
the axion solution. The history of axion is\\
\indent ---1997: The Peccei-Quinn symmetry~\cite{pq} 
and PQWW axion~\cite{ww}\\
\indent ---1978: No PQWW axion~\cite{tokyo} and 
calculable models~\cite{cal}\\
\indent ---1979: Invisible axion~\cite{kim}\\
\indent ---1984: Superstring axion~\cite{witten}\\
\indent ---1985: Composite axion~\cite{comp}\\
\indent ---1988: Anomalous $U(1)$ axion~\cite{anom}

These axion models introduce a pseudoscalar field $a$ in the
effective Lagrangian. One can show that $\bar\theta=0$ is the
minimum of the free energy from
the simple and elegant argument regarding the pseudoscalar
nature of the $F\tilde F$ term~\cite{vw}. After integrating
out the quark fields, one obtains the following path integral
in the Euclidian space,
\begin{equation}
\int [dA_\mu]\prod_i{\rm Det}(D^\mu\gamma_\mu+m_i)e^{-\int
d^4x({1\over 4g^2}F^2-i\bar\theta\{F\tilde F\})}
\end{equation}
where \{ \} includes the factor $1/32\pi^2$. Note that the
$\bar\theta$ term is pure imaginary in the Euclidian space.
In the Euclidian space, ${\rm Det}(D^\mu\gamma_\mu+m_i)>0$.
Therefore, defining the integral as a function of $\bar\theta$,
we obtain the following inequality, due to the Schwarz inequality,
\begin{eqnarray} 
&e^{-\int d^4x V[\bar\theta]}\equiv \int [dA_\mu]\prod_i{\rm Det}
(D^\mu\gamma_\mu+m_i)e^{-\int d^4x({1\over 4g^2}F^2-i\bar\theta
\{F\tilde F\})}\nonumber\\
&\le \int[dA_\mu]\left|\prod_i{\rm Det}(D^\mu\gamma_\mu+m_i)
e^{-\int d^4x({1\over 4g^2}F^2-i\bar\theta\{F\tilde F\})}\right|\\
&=e^{-\int d^4xV[0]}\nonumber
\end{eqnarray}
which implies
\begin{equation}
V[\bar\theta]\ge V[0].
\end{equation}

Thus we obtain that $\bar\theta=<a>/F_a=0$ is the minimum
of the axion potential.
In the above proof, we neglected the weak CP violation. Inclusion
of the weak CP violation shifts the position of the minimum but
not very much~\cite{gr}.

This potential is almost flat for a large $F_a$. Thus the classical
axion field starts to oscillate very late, $T\sim 1$ GeV, 
measured at the scale of $F_a$,
which leads to the significant cold axion energy density in the
universe~\cite{pww}.

\section{Axino-gravitino Cosmology}

Supersymmetrization of axion introduces $s$ (saxion, the scalar
partner of $a$) and $\tilde a$ (axino, the fermionic partner of
$a$). Mass of $s$ is of order $M_{SUSY}$. But the axino mass
can be lighter. It depends on models \cite{chun}. With the presence
of the axino, it is important to know what is the LSP with the
unbroken $R$--parity. The cosmological scenario with
($\tilde a=$ LSP) has been studied extensively by Rajagopal
{\it et al.}\cite{rtw}.

The case with (gravitino = LSP) arises in no-scale supergravity
models~\cite{noscale} and in the gauge mediated SUSY breaking 
scenario~\cite{dn}.
Phenomenological applications of the gravitino LSP are the
axino-gravitino cosmology~\cite{ckk} and 
the anomalous $\gamma$ event of the
CDF group. The axion--axino--gravitino coupling is given by
\begin{equation}
{\cal L}_{a\tilde a\tilde G}={1\over M_P}\bar\psi_\mu
\gamma^\nu\partial_\nu z^*\gamma^\mu\tilde a_L+{\rm h.c.}
\end{equation}
where $z=(s+ia)/\sqrt{2}$ and $M_P\simeq 2.44\times 10^{18}$ GeV. 
For the light gravitino, the Goldstino
component $\xi$ dominates and replacing $\psi_\mu=i\sqrt{2/3}
(1/m_{3/2})\partial_\mu\xi$ gives the lifetime of axino
from the $\tilde a\rightarrow \tilde G+a$  decay as \cite{ckk,kk}
\begin{equation}
\tau_{\tilde a}={96\pi M_P^2m_{3/2}^2\over m^5_{\tilde a}}\simeq
1.2\times 10^{12}\left({{\rm MeV}\over m_{\tilde a}}\right)^5
\left({m_{3/2}\over {\rm eV}}\right)^2\ {\rm sec}.
\end{equation}
The decoupling temperature of $\tilde a$ is \cite{rtw}
\begin{equation}
T_{\tilde a}=10^{11}\left({F_a\over 10^{12}{\rm GeV}}\right)^2
\left({0.1\over \alpha_c}\right)^3\ {\rm GeV}.
\end{equation}
So it is interesting to note that the MeV axino mass with
low energy SUSY breaking leads to the axino lifetime around
the time of galaxy formation. This axino may affect the
formation of large scale structures.

\subsection{Cosmology with late decaying particles}

The cold dark matter (CDM) was successful before the COBE data.
This assumes a flat universe with 5--10 \% baryonic dark matter
and the rest CDM. The seed fluctuations with the inflationary
idea are assumed to be of the scale invariant form. In this case
the evolved spectrum is given by \cite{bardeen}
\begin{equation}
|\delta_k|^2={Ak\over (1+\alpha k+\beta k^{3/2}+\gamma k^2)^2}
\end{equation}
where $A$ is a normalization constant, $\alpha=1.7 l$, $\beta=9.0 
l^{3/2}$, $\gamma=1.0 l^2$, and
\begin{equation}
l=(\Omega h^2)^{-1}\theta^{1/2}\ {\rm Mpc}.
\end{equation}
Here $\theta=\rho_{\rm rel}/1.68\rho_\gamma$ measuring the present
energy density of all relativistic particles relative to those of
photons and neutrinos. The COBE data fixed the normalization, and
the CDM model needed modifications.

\begin{figure}
\centerline{\epsfig{figure=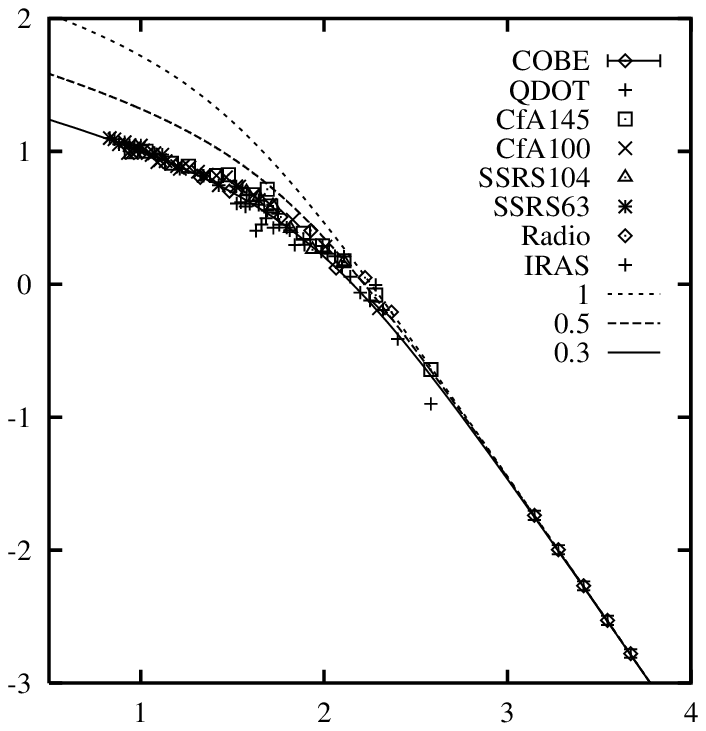,height=2.5in}}
\caption{log$_{10}{\delta\rho\over\rho}$ versus
log$_{10}{\lambda\over h^{-1}{\rm Mpc}}$.\label{fig1}}
\end{figure}

Successful fits are: (i) $\Omega_\Lambda\simeq 0.8$ and 
$\Omega_{\rm CDM}\simeq 0.2$, (ii) $\Omega_{\rm CDM}\simeq 0.7$
and $\Omega_{\rm HDM}\simeq 0.3$, and (iii) $\Omega_{\rm CDM}
=0.2-0.3$. One can mimick Case (iii) even in the $\Omega=1$
universe if $\theta\ne 1$. In effect, it resembles a CDM+HDM 
universe ($\equiv$ MDM). 
The $\theta>1$ universe is obtained in cosmology with
late decaying particles. The decay products must be relativistic
and noninteracting to mimick Case (iii). The late decaying
particle cosmology was first considered by Bardeen, Bond and Efstathiou
\cite{bardeen}, then applied for the by-now dead 17 keV neutrino
by Bond and Efstathiou \cite{bond}. After the COBE data, it was
first considered by Chun {\it et al.}~\cite{ckk}, and
this idea was later applied to $\nu_\tau$ by others~\cite{others}. 
For the structure formation,
the radiation-matter equality point $R_{EQ}$ is important.
At $R_{EQ}$ the structure scale $\lambda_{EQ}$ is given by
\begin{equation}
\lambda_{EQ}\simeq 30(\Omega h^2)^{-1}\theta^{1/2}\ {\rm Mpc}.
\end{equation}

\subsection{Axino-gravitino cosmology}

A light axino decays to an axion and a gravitino via the
interaction given in Eq. (7). Normally, one would expect
a coupling supressed by $M_P$, but the Goldstino component
dominates whose coupling is supressed by $F_S$. Namely,
the gravitino coupling is of the form $(1/F_S)(\partial_\mu\xi)
J^\mu$ where $J^\mu$ is the supercurrent. In this case, the
axino lifetime is given in Eq. (8). The detail energy densities
of respective species are given in Fig. 2.

\begin{figure}
\centerline{\epsfig{figure=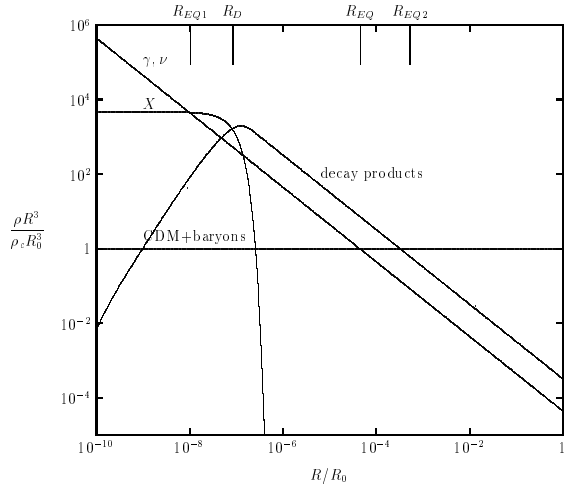,height=2.5in}}
\caption{Scale factor versus energy densities of particles. \label{fig2}}
\end{figure}

When the cosmic scale factor exceeds $R_{EQ1}$, $\tilde a$
dominates the mass density of the universe. The cold axion
dominates the energy density of the universe after the
scale factor exceeds $R_{EQ2}$. In Fig. 3, $R_{EQ}$ is the radiation-
matter equality point in the CDM model. Thus, the axino-gravitino
cosmology extends the time for $R_{EQ}$ by a factor
\begin{equation} 
{R_{EQ2}\over R_{EQ}}\simeq 1+\left({\tau_{\tilde a}\over 
t_{EQ1}}\right)^{2/3}
\end{equation}
compared to the CDM model. Roughly, the fluctuation spectrum is
characterized by two length scales,
\begin{eqnarray}
&\lambda_{EQ1}\simeq 8\times 10^{-2}\left({{\rm MeV}
\over m_{\tilde a}Y}\right) \nonumber\\
&\lambda_{EQ2}\simeq 30(\Omega h^2)^{-1}\left[
1+\{{1\over 0.55}({\tau_{\tilde a}\over {\rm sec}})({m_{\tilde a}Y
\over {\rm MeV}})^2\}^{2/3}\right]^{1/2}
\end{eqnarray}
kpc where $Y=n_{\tilde a}(T)/s(T)$. $\lambda_{EQ2}$ corresponds to
the size of galaxies, and it is interesting if $\lambda_{EQ1}$
corresponds to the size of globular clusters.
The condition that the axino model mimicks the mixed dark matter
model in the $\Omega=1$ universe is
\begin{equation}
\left({\tau_{\tilde a}\over {\rm sec}}\right)
\left({m_{\tilde a}Y\over {\rm MeV}}\right)^2\simeq 0.55\left[
({h\over 0.2})^2-1\right]^{3/2}.
\end{equation}
The nucleosynthesis bound is
\begin{equation}
\left({m_{\tilde a}Y\over {\rm MeV}}\right) < 0.107 \ \ 
{\rm for}\ \ m_{\tilde a}>1\ {\rm MeV}.
\end{equation}
The energy density bound is
\begin{equation}
\left({\tau_{\tilde a}\over {\rm sec}}\right)\left(
{m_{\tilde a}Y\over {\rm MeV}}\right)<2\times 10^6h^3.
\end{equation}
These conditions are shown in Fig. 3.

\begin{figure}
\centerline{\epsfig{figure=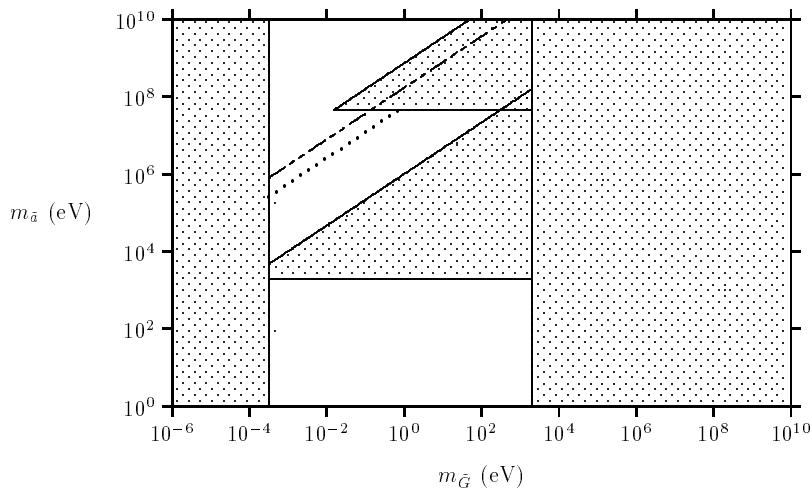,height=2.5in}}
\caption{Exclusion plot in the axino and gravitino mass plane. 
 \label{fig3}}
\end{figure}

The size $\lambda_{EQ1}$ depends on the reheating temperature
$T_R$. If $T_R\gg T_{\tilde a}$, then one obtains 
$\lambda_{EQ1}\sim 44 ({\rm MeV}/m_{\tilde a})$ kpc. If
$T_R\ll T_{\tilde a}$, then $\lambda_{EQ1}\sim
11 (F_a/10^{12}\ {\rm GeV})^2(10^6\ {\rm GeV}/T_R)(
{\rm GeV}/m_{\tilde a})^2$ kpc. The dotted line in Fig. 3
is Eq. (15).

\section{Conclusion}

The simultaneous solution of the strong CP problem and
the gauge hierarchy problem leads to axino and
gravitino. The LSP is probably the axino or
the gravitino. If the gravitino is the LSP, its effect
on the large scale structure formation can mimick the
MDM for appropriate mass parameters of
the axino and gravitino. And there may be smaller 
size ($\sim$ tens of kpc) structures also.

\section*{Acknowledgments}
This work is supported in part by Distinguished Scholar
Exchange Program of the Korea Research Foundation, 
by the Hoam Foundation and by NSF-PHY 92-18167.


\begin{thebibliography}{99}

\bibitem{ckk}E. J. Chun, H. B. Kim and J. E. Kim, \Journal{\PRL}{72}
{1956}{1994}.

\bibitem{kk}H. B. Kim and J. E. Kim, \Journal{\NPB}{433}{421}{1995}.

\bibitem{bpt}A. A. Belavin, A. Polyakov, A. Schwartz, and Y. 
Tyupkin, \Journal{\PLB}{59}{85}{1974}.

\bibitem{altarev}I. S. Altarev {\it et al.}, \Journal{\PLB}{276}{242}{1992}.

\bibitem{pq}R. D. Peccei and H. R. Quinn, \Journal{\PRL}{38}{1440}{1977}.

\bibitem{ww}S. Weinberg, \Journal{\PRL}{40}{223}{1978}; 
F. Wilczek, \Journal{\PRL}{40}{279}{1978}. 

\bibitem{tokyo}R. D. Peccei in {\it Proc. 19th Int. Conf. High Energy
Physics (Aug., 1978)}, ed. S. Homma {\it et al.} (Phys. Soc. of Japan,
Tokyo, 1979), p.385.

\bibitem{cal}M. A. B. Beg and H.-S. Tsao, \Journal{\PRL}{41}{278}{1978}.

\bibitem{kim}J. E. Kim, \Journal{\PRL}{43}{103}{1979}.

\bibitem{witten}E. Witten, \Journal{\PLB}{149}{351}{1985}

\bibitem{comp}J. E. Kim, \Journal{\PRD}{31}{1733}{1985}.

\bibitem{anom}J. E. Kim, \Journal{\PLB}{207}{434}{1988}.

\bibitem{vw}C. Vafa and E. Witten, \Journal{\PRL}{53}{535}{1983}.

\bibitem{gr}H. Georgi and L. Randall, \Journal{\NPB}{276}{241}{1986}.

\bibitem{pww}J. Preskill, M. B. Wise and F. Wilczek, 
\Journal{\PLB}{120}{127}{1983}; L. F. Abbott and P. Sikivie, 
\Journal{\PLB}{120}{133}{1983};
M. Dine and W. Fischler, \Journal{\PLB}{120}{137}{1983}.

\bibitem{chun}E. J. Chun, J. E. Kim and H. P. Nilles, 
\Journal{\PLB}{287}{123}{1992}; E. J. Chun and A. Lukas,
\Journal{\PLB}{357}{43}{1995}.

\bibitem{rtw}K. Rajagopal, M. S. Turner and F. Wilczek,
\Journal{\NPB}{358}{447}{1991}.

\bibitem{noscale}J. Ellis, A. B. Lahanas, D. V. Nanopoulos, 
and K. Tamvakis, \Journal{\PLB}{134}{429}{1984}.

\bibitem{dn}M. Dine and A. E. Nelson, \Journal{PRD}{48}{1277}{1993}.

\bibitem{bardeen} J. M. Bardeen, J. R. Bond and G. Efstathiou,
{\it Astrophys. J.} {\bf 321}, 28 (1987).

\bibitem{bond}J. R. Bond and G. Efstathiou, 
\Journal{\PLB}{265}{245}{1991}.

\bibitem{others}M. S. Turner, \Journal{\PRL}{72}{3754}{1994};
R. N. Mohapatra and A. Riotto, \Journal{\PRL}{73}{1324}{1994};
A. D. Dolgov, S. Pastor and J. W. Valle, preprint astro-ph/9506011.

\end{thebibliography}
\end{document}